\documentstyle[aps,preprint,amssymb]{revtex}
\draft
\title{A model of cation ordering in A(B'$_{x}$B''$_{1-x}$)O$_3$ relaxors}

\author{
S.\ Lapinskas} 
\address{
Faculty of Physics, Vilnius University, Saul\.{e}tekio 9, 2054 Vilnius,
Lithuania and Department of Theoretical Physics, The Royal Institute of Technology, S--100 44 Stockholm, Sweden}
\author{
E.\ E.\ Tornau}
\address{
Semiconductor Physics Institute, Go\v{s}tauto 11, 2600 Vilnius, Lithuania}
\author{
A.\ Rosengren}
\address{
Department of Theoretical Physics, The Royal Institute of Technology,
S--100 44 Stockholm, Sweden}

\begin{document}
\maketitle
\begin{abstract}
We have shown that the lattice-gas model with four repulsive {\it
pair} interaction constants (corresponding to the four nearest coordination
shells) on a simple cubic lattice is a sufficient model to
describe the main types of cation ordering in relaxors. The phase diagram,
obtained by the cluster variation method, shows the sequences of
transitions between the phases $1:2\rightarrow1:1\rightarrow$
disordered and $1:2\rightarrow$ disordered. 

\end{abstract}

\section{INTRODUCTION}

Ferroelectric and piezoelectric properties of so-called relaxors,
i. e. some complex mixed-metal perovskites with general chemical
formulae A(B'$_{x}$B''$_{1-x}$)O$_3$, depend on the state of order in
the (B', B'') sublattice~\cite{rand27,bell18,burt87}. It is well-known
that such relaxor materials as Pb(B'$_{1/2}$Nb$_{1/2}$)O$_3$ (B'=In and
Yb) are ferroelectric when disordered and antiferroelectric when
ordered~\cite{isup75}. The correlation between the order-disorder tendency
of (B', B'') ions and the  electromechanical ($e/m$) coupling factor, an
important parameter for fabrication of high-quality piezoelectric
materials, is also known~\cite{yama88}. Thus the compositional
ordering in the (B', B'') sublattice might be essential for the different
applications in which relaxor materials are nowadays used: the
A=Pb-based relaxors are widely used in piezoelectric transducers for
mostly medical diagnostic application, in actuators, capacitors,
ultrasound and sonar listening devices, and the A=Ba-based relaxors are
used in production of dielectric resonators and other high-frequency
applications.

Apart from well-known examples, such as BaTiO$_3$ with occupation
of the (B', B'') sublattice exclusively by 4-valent Ti ions ($x=0$), 
there are two
other main stoichiometries of this sublattice with  $x=1/2$ and
$x=1/3$. While tetravalent (B', B'') ions show no ordering
(e. g. Pb(Zr$_x$Ti$_{1-x}$)O$_3$~\cite{cros05}), samples made of
heterovalent ions, i.e. 3-valent B'= In, Sc, Yb and 5-valent B''= Nb,
Ta or 2-valent B'= Zn, Mg, Ni, Co and 6-valent B''= W, Mo, in most
cases exhibit long range order below an ordering transition temperature
for $x=1/2$. The same is true for most of the 2-valent B' and 5-valent B''
ions at $x=1/3$ (for experimental data on ordering types
and temperatures see Tables in Refs~\cite{burt87,burt42}).

Basically, two types of (B', B'') sublattice ordering are found experimentally.
For $x=1/2$ compounds the 1:1 stacking of B' and B'' layers perpendicular to the
[111] axis of the simple cubic cell is observed in all A=Ba- and
Pb-based A(B'$_{x}$B''$_{1-x}$)O$_3$ materials. For $x=1/3$ the 1:2
structure with stacking of B', B'', B'' layers is found in all
A=Ba-based relaxors with B'=Zn, Mg and B''=Nb, Ta. This type of
ordering is the ground state for $x=1/3$ in a Coulomb-type of B'-B''
interaction model~\cite{bell18}. However the same four materials with
Ba fully substituted by Pb in the A sublattice demonstrate either
short-range 1:1 ordering (e.g. Pb(Zn$_{1/3}$Nb$_{2/3}$)O$_3$ (PZN),
Pb(Mg$_{1/3}$Nb$_{2/3}$)O$_3$ (PMN)) or long-range 1:1 ordering
(Pb(Mg$_{1/3}$Ta$_{2/3}$)O$_3$ (PMT)). This structure might be viewed
as an alteration of B', B'' sequence along the [111] axis with the B''
site occupied by Ta(Nb) and B' by a disordered mixture of
2/3Mg(Zn)+1/3Nb(Ta)~\cite{burt87}. Especially interesting is the type
of (B', B'') ordering observed in so-called relaxor ferroelectric
alloys, for example $(1-y)$Ba(Zn$_{1/3}$Ta$_{2/3}$)O$_3$+$y$BaZrO$_3$
(BZT-BZ) or $(1-y)$Ba(Mg$_{1/3}$Nb$_{2/3}$)O$_3$+$y$BaZrO$_3$
(BMN-BZ). For small values of $y<0.02$ the 1:2 ordering is observed. At
intermediate values ($0.04<y<0.25$ for BZT-BZ and $0.05<y<0.15$ for
BMN-BZ) the 1:1 ordering is observed in small microregions in a
disordered matrix. At even higher $y$ disordered structures
are observed.

Therefore it is interesting and important to find the appropriate model and
from this model obtain the phase diagram (PD)
which would account for known experimental facts including
the disordered phase at high temperatures and the 1:1 and 1:2 phases at their
corresponding stoichiometries. This problem is closely related to
the problem of finding a minimal set of interaction constants
in an Ising-type model for a simple cubic lattice.

One of the first models, proposed to simulate the images found in
HRTEM experiments for PMN, accounted for the nearest neighbor (NN) and
next-nearest neighbor (NNN) interactions between (B', B'') ions
supplemented by an electrostatic (Coulomb) energy term
$E_{el}$~\cite{burs01}. At $E_{el}=0$ this Monte Carlo simulation
yields the separation of initial 1:2 system into locally ordered
negative 1:1 (Pb(Mg$_{1/2}$Nb$_{1/2}$)O$_3$)$^{1-}$ domains surrounded
by positive regions of (PbNbO$_3$)$^{1+}$. The minimization of
electrostatic energy between these domains, realized by the inclusion
of $E_{el}$, leads to much smaller 1:1 regions percolating the
lattice. The obtained texture, which gives the superlattice
reflections much closer to those seen in experiments, might be
interpreted as a vanishing of the two-phase segregation. Another Monte
Carlo simulation~\cite{bell18}, used the Coulomb interaction as the
only type of interaction energy present in an ideal system of ionic
charges. The comparison of the results of this model to different
types of long-range structures in relaxors with $x=1/2$ and $x=1/3$
supported very much the viewpoint that the Coulomb interaction between
ions might be the driving force of the ordering.

There are also known several studies of the ground states of a simple
cubic lattice using the Ising model with just few short-range
interaction constants taken into
account~\cite{kats26,kabu30,lipk18}. However, the range of interaction
of these models was not sufficient for the occurence of the ground
states found in relaxors, the 1:2 phase in particular. Namely, the
number of interaction constants were limited either to those three
corresponding to the three main distances of the cube (i. e. NN, NNN
and third-nearest neighbor (3NN))~\cite{kats26,kabu30} or to a very
particular set of NN, NNN and fourth-nearest neighbour (4NN)
interactions as well as to the smallest triangle interaction and
linear triplet, but omitting 3NN interaction~\cite{lipk18}. Therefore
much more important to the studies of stability of ordered phases in
relaxors were the works of Burton~\cite{burt87} and McCormack and
Burton~\cite{mcco53} where the stability conditions for the 1:2 phase
were investigated in particular.  These authors have concluded that
using simple pair interactions in the Ising-type Hamiltonian up to
fourth-nearest neighbors is not sufficient to obtain the
experimentally observed superstructures as ground states.  They
obtained the PD with phase transition sequences
$1:2\rightarrow1:1\rightarrow$ disordered and $1:2\rightarrow$
disordered using many-body interactions such as linear triplet and,
so-called, cube-222. However the values of these interaction
constants, chosen~\cite{burt87} for the calculation of the PD, cannot
be reconciled with the relative energies for several of the most
probable phases obtained from first-principles calculation by the same
authors~\cite{mcco53}.

In this work we present the PD of a lattice-gas model
with four repulsive {\it pair} interaction constants (NN, NNN, 3NN and 4NN)
taken into account in a simple cubic lattice. Our calculations,
performed by the cluster variation method (CVM), show that this model 
is sufficient to explain the main ordering features of relaxors,
and that no multi-point interactions are really necessary.

\section{MODEL AND PHASE DIAGRAMS}

We consider a lattice-gas model which is described by the Hamiltonian
\begin{equation}
{\cal H} = \sum_{i<j} v_{ij}n_in_j - \mu \sum_{i} n_i, \label{hamil}
\end{equation}
where $n_i = 1$ when the site is occupied by the atom B' and 0 when it
is occupied by B''. It can be shown \cite{ducast} that the effective interaction
constant $v_{ij}=v_{ij}^{B'B'}+v_{ij}^{B''B''}-2v_{ij}^{B'B''}$, where
$v_{ij}^{MN}$ is the interaction energy between the atoms $M$ and $N$,
and $\mu=\mu_{B'}-\mu_{B''}+v^{B''B''}-v^{B'B''}$
($v^{MN}=\sum_{i}v_{ij}^{MN}$) is the chemical potential of the system
expressed via the chemical potentials of the different cations.
The lattice gas model can be
mapped to the more common Ising model in an external field via the
relation between $n_i$ and the Ising spin variables $s_i = \pm 1$,
$s_i = 2n_i-1$, which implies the simple relation $v_{ij} = 4J_{ij}$
between the Ising interaction constants $J_{ij}$ and $v_{ij}$.

We are going to show that repulsive pair interactions in
the four nearest coordination shells are sufficient to describe the
experimentally observed 1:1 and 1:2 phases and the phase transitions
between these phases. This statement contradicts earlier ground-state 
analysis~\cite{mcco53}, thus the Appendix is devoted to show 
that our approach is justified. Here we just want to give a simple
indirect argument supporting our choice. 

According to the ground-state analysis~\cite{kabu30} for repulsive
$v_1, v_2, v_3 > 0, v_4=0$ and the case $v_3 < v_2/2$, $v_2 <
v_1/4+v_3$, four ordered ground-states take place at $x=\langle n_i
\rangle=1/2, 3/8, 1/4$ and 1/8. The ground-state structure is
$S(0,6,0;{1\over 2})$ or 1:1 at $x=1/2$, $S(0,4,0,{3\over 8})$
(hereafter labelled as 3/8) at $x=3/8$, $S(0,0,4,{1\over 4})$ (1/4) at
$x=1/4$ and $S(0,0,0;{1\over 8})$ (1/8) at $x=1/8$. The structure
notations here, $S(p_1, p_2, p_3; x)$, are taken from
Ref~\cite{kabu30}, and $p_i$ means the number of B'-B' bonds per B'
atom in the $i$-th coordination shell of the B'$_x$B''$_{1-x}$
structure ($i$=1, 2, 3 corresponds to NN, NNN and 3NN respectively).
The phases are shown schematically in Fig.\ref{fig0}.  The 1/4 phase is $bcc$
with a size of a new unit cell $a=2a_0$, the 1/8 phase is $sc$ with
$a=2a_0$, and the 3/8 phase could be regarded as 1:1 phase with
substracted 1/8 structure (atoms are removed from the sites which are
part of $a=2a_0$ cubic structure). Here $a_0$ is the lattice parameter
of the initial $sc$ unit cell. The ground-state at $x=1/3$ (the
stoichiometric value for the 1:2 structure) is a mixture of 1/4 and
3/8 structures taken with proper weights. One can easily check that
for $v_4=0$ the energy of the 1:2 structure at $x=1/3$ is exactly the
same as that of the mixture, i.e. the 1:2 phase is just marginally
unstable. By introducing an arbitrarily small $v_4 > 0$ one can make
the 1:2 structure stable albeit only at very low temperatures. This is
because the energy of the mixture increases with $v_4$, but the energy
of the 1:2 phase is independent of $v_4$.

When choosing the empiric values of the interaction constants we assume $v_1
= 1$, since the negative value would cause decomposition rather than
ordering and the absolute value just gives the energy scale which can
always be adjusted to the experiment. It would be interesting to
explore all possible situations in the remaining three-dimensional
space of $v_2, v_3$ and $v_4$ for which the ground state is the 1:2
structure at $x=1/3$. This is too ambitious however, since not much is
known about the PDs for the simple cubic lattice even for $v_4 = 0$.
Thus to simplify our task, we base our choice on the results of
Ref.~\cite{bell18}, that the interactions might be of
electrostatic nature (similar atoms tend to repel each other) and take
{\it all} $v_i$ to be positive.

The calculations of the PDs have been performed using the
CVM~\cite{kiku88}, a powerful variational method to treat phase transition
problems. In this method the local interactions between a particular
group of spins (cluster) are exactly taken into account. The
interaction of the cluster spins with those being outside the cluster
are accounted for by the, so-called, effective fields which are the
variational parameters in the version~\cite{zubk53} of the CVM used
here. The PD for $v_2=0.2, v_3 = 0.06$ and $v_4 = 0$ obtained using
the 8+7 - point (cube + octahedron) approximation of the CVM is presented in
Fig.~\ref{fig1}. The values of interactions were chosen to obtain the 1:1
phase down to $x < 1/3$ at intermediate temperatures (which
are below the maximum of the 1:1 phase at $x=1/3$, but above all other
ordered phases). This PD could describe those
A(B'$_{1/3}$B''$_{2/3}$)O$_3$ materials, which do not exhibit 1:2 ordering,
but the ground state in this case would be a mixture of 1/4
and 3/8 phases. By setting $v_2 = v_3 = 0$ we could
make other structures disappear arriving at the classical
antiferromagnetic Ising result with only 1:1 ordering. As was already
mentioned, there is no 1:2 phase in this PD, since at $x=1/3$ the
energy of the 1:2 structure is equal to that of the proportional mixture
of 1/4 and 3/8 phases.

For $v_4 > 0$ the 1:2 phase becomes energetically favourable with
respect to the mixture of neighboring phases. The higher $v_4$, the
higher temperatures the 1:2 phase would reach. Other consequences of
$v_4 > 0$ on the PD are as follows. First, the domain of the 1:1 phase
at intermediate temperatures should shrink at higher $x$, since the
inclusion of $v_4$ increases the energy of the 1:1 phase.  Second, the
degeneracy of the 1/8 and 3/8 phases is removed at $v_4 > 0$ (at
$v_4=0$ any [100] or equivalent plane of the 1/8 phase can be moved
along any vector in that plane and the depleted [100] planes of the
3/8 phase can be moved along $v_2$ in that plane without change of
energy).  Actually, for small $v_4 > 0$ the ground state at $x=3/8$ is a
new 3/8 phase (3/8'), which is obtained from the ``original'' 3/8
phase by shifting every second depleted [100] plane along the (011)
direction (see Fig.\ref{fig0}). Also at $x=1/8$ the ground state for
$v_4 > 0$ is different from the ``original'' 1/8 phase (we do not
examine this phase as it has no relevance to the vicinity of
$x=1/3$). Third, a number of ordered structures with longer period
should appear below $x=1/8$ and above $x=3/8$, which we also do not
consider here. For small $v_4 > 0$, the transition from
1:2 to 3/8' phase takes place between $x=1/3$ and $x=3/8$. Another
transition from 1/4 to 1:2 occurs between $x=1/4$ and $x=1/3$.  The PD
for $v_2=0.2, v_3 = 0.06$ and $v_4 = 0.02$ is presented in
Fig.\ref{fig2}. The PD is not complete as the phases to the left of the
1/4 phase and to the right of the 3/8' phase are not considered. We
obtain a rather narrow 1:2 phase in ($T, x$) coordinates. Meanwhile in
the ($T, \mu$) diagram, there is quite a large domain of chemical
potential corresponding to the 1:2 structure. It should be also noted
that for $v_4=0.02$ we could not obtain the 3/8' phase by the CVM,
since the mixture of the 1:2 and off-stoichiometric 1:1 phase has lower
energy. We calculated the high-temperature metastability limit of the
3/8' phase by Monte Carlo simulations only (see Fig.\ref{fig2}b).
Still we can conclude that if the 3/8' phase exists, it exists
only below $T\approx 0.08$.

For $2.01<\mu<2.06$ and $0.13<T<0.25$ the 8+7 - point approximation of
the CVM has unexpectedly shown some instability resulting in the gap
on the phase boundary between 1:1 and 1:2 phases and the strange
reentrance of the second order transition line between disordered and
1:1 phase. It is interesting to note, that reducing the variational
space of the CVM by omitting all the correlations with more than two
sites, the problems related to the second order phase transition line
disordered - 1:1 disappear (dotted line in Fig.\ref{fig2}), but those
related to the transition 1:1 - 1:2 become worse. Such a behavior of
the 8+7 - point approximation is clearly related to the increased
range of interactions which can not be properly accommodated within
the cube and octahedron clusters. It seems that these problems appear
at any $v_4>0$ (we checked the values between 0.01 and 0.04, also
slightly varying $v_2$ and $v_3$). At $v_4 = 0$ we had no similar
problems with the 8+7 - point approximation. To be sure, we have
performed calculations with a 27 - point ($2a_0\times2a_0\times2a_0$
cube) approximation of the CVM. Besides the increase of accuracy due
to larger basic cluster (a general feature of the CVM), the 27-point
cluster has an advantage of including all interactions into single
basic cluster, thus avoiding occasional breakdowns, sometimes observed
when basic clusters are too small. Unfortunately, the minimisation of
the free energy then becomes more complicated, since the number of
variational parameters drastically increases. Since it is not feasible
to treat all $2^{27}\approx 134$ mln configurations of the 27 - point
cluster, we excluded the possibility for the NN sites to be occupied
($v_1=\infty$). This NN exclusion makes a big difference at high
temperatures, as the 1:1 phase is always present above some critical
concentration. However we expect only small changes at low
temperatures, where 1:2 and the rest of ordered phases exist, since
due to $v_1\gg T$ the probability of the NN being occupied is very
low, especially for small concentrations. The NN exclusion decreases
the number of configurations of the 27 - point cluster down to 70663,
which is tractable with some effort. The results of the 27 - point CVM
are presented by dashed lines in Figs.~\ref{fig1} and
\ref{fig2}. Indeed, while there are large discrepancies between the
results obtained by the 8+7- and 27 - point approximations at higher
temperatures due to the exclusion, at low temperatures the lines are
pretty close. It is rather difficult to compare accuracies of the two
approximations, since formally they describe different models. The 27
- point approximation does not suffer from the problems of the 8+7 -
point one and confirms its results by filling the gap in the PD. To be
sure, we have also applied the NN exclusion to the 8+7 - point
approximation (to check whether the exclusion itself could hide the
problems with the CVM), but the problems remained.

To support our CVM calculations, we have performed some limited Monte
Carlo simulations using a simple Metropolis algorithm on a $24\times
24\times 24$ lattice with periodic boundary conditions and around
50000 sweeps for each point. These points are plotted in
Fig.\ref{fig2} with circles. First-order transitions are difficult to
locate accurately by the Monte Carlo method, so the points on the
first-order phase transition lines are actually metastability limits
of the corresponding low-temperature phase, a true phase transition
should be slightly below. It is much easier to find metastability
limits, if one starts from a perfectly ordered structure at low
temperatures and gradually increases the temperature until the order
changes or is destroyed.  We show in Fig.\ref{fig3}, that the
metastability limits obtained by the CVM and Monte Carlo calculations
are quite close, thus we expect the location of the true first-order
phase transition to be correct too.  It is seen that these results
nicely confirms our CVM data.

In summary, a lattice gas model for a simple cubic lattice with four
nearest repulsive pair interaction constants was proposed.  The presented
PD posesses all features of cations ordering observed in
relaxor compounds, namely, a phase transition sequence $1:2\rightarrow
1:1\rightarrow$ disordered at $x=1/3$. The increase of $v_4$ from 0.02 to 0.03
shifts the phase boundary between disordered and 1:1 phases to the
region $x>1/3$, thus resulting at $x=1/3$ in a phase transition
$1:2\rightarrow$ disordered, which was also observed in a number of
relaxor materials. 
Another interesting feature of the PD is that a
pure 1:2 phase reaches its stoichiometric concentration only at $T=0$,
while at any finite temperature it separates into a mixture of the 1:2
phase with a slightly lower concentration and a small amount of
adjacent phase (e.g. 1:1). This indicates that a slightest increase
in concentration above stoichiometric value tends to destroy the 1:2
ordering, making it favorable for the excess ions to form islands of
another higher concentration phases. Could this be an explanation of
the nanoregions, which are observed in some
A(B'$_{1/3}$B''$_{2/3}$)O$_3$ compounds and, most likely, are
responsible for the relaxor properties, besides the electrostatic
arguments?

This work was supported by The Swedish Academy of Sciences, The
Swedish Institute, The Swedish Natural Research Council and SSF 
({\it Swedish Foundation for Strategic Research}). SL is
indebted to The Royal Institute of Technology, Stockholm for kind
hospitality.

\appendix
\section{}
Here we apply the partial vertex enumeration (PVE) 
procedure~\cite{mcco53}
for interaction clusters to our four two-point clusters corresponding to
NN, NNN, 3NN and 4NN pairs.
The PVE technique for the ground state problem is based up on the more
general vertex enumeration of the configuration
polyhedron method~\cite{ducast}. It is easier described in terms of 
the Ising model and the
outline is as follows.  Consider a cluster $\alpha$ of $|\alpha|$
lattice sites which includes all interactions of the Hamiltonian. It
is shown~\cite{ducast} that the probability function
$\rho_{\alpha}(\{s\})$ for the configuration $\{s\}$ (the set of Ising
spins $s_i$ at every site $i$) of the cluster $\alpha$ is a linear
function of the spin products $s_{\beta} = \prod_{i\in\beta}s_i$ of
all subclusters $\beta$ of the cluster
\begin{equation}
\rho_{\alpha} = {1\over {2^{|\alpha|}}}(1 +
\sum_{\beta\subset\alpha,\beta\neq\o} \nu_{\beta}s_{\beta}), \label{prob1}
\end{equation}
where $\nu_{\beta} = \langle s_{\beta}\rangle$ are cluster correlation
functions.
The requirement $0\leqslant\rho \leqslant 1$ imposes constraints for
the allowed values of $\nu_{\beta}$, and equations
\begin{equation}
\rho_{\alpha}(\{s\}) = 0 \label{prob2}
\end{equation}
for each symmetry distinct configuration $\{s\}$ define hyperplanes in
the space spanned by $\nu_{\beta}$. These hyperplanes are boundaries in
a space of allowed values of $\nu_{\beta}$ which is a configurational
polyhedron (CP). The possible ground states of the model reside in the
vertices of the CP, because the energy is a linear function of
$\nu_{\beta}$, and the extremum of the linear function is reached on
the boundary of its domain of definition. Since not all $\nu_{\beta}$
necessarily contribute to the energy $E = \langle{\cal H}\rangle$ ,
one needs to project the CP on the subspace of those $\nu_{\beta}$,
which contribute to the energy.

However it may be impractical to make a full vertex enumeration for
large clusters, when interactions beyond the 3NN are
considered.  Therefore the PVE is used (see Ref~\cite{mcco53}) to
determine whether in the space of $\nu_{\beta} (\beta\subset\alpha$)
included in a certain cluster $\alpha$ (or several clusters), the
point, corresponding to the given structure, is a vertex of the CP.
It is a vertex if the number of faces it belongs to is not less than
the dimensionality of the space (number of distinct $\nu_{\beta}$).
Then for a given (periodic) structure one obtains values of
$\nu_{\beta}$ by averaging over the unit cell of the structure, and
checks how many equations (\ref{prob2}) are satisfied by these
values. It is found~\cite{mcco53} that at least some
many-body interactions in the cube and a linear triplet are necessary
for the 1:2 phase to be a vertex.

Consider PVE for our four pair clusters ($\alpha$ = 1,2,3,4).
The configuration probabilities for each cluster are
$\rho_{\alpha} = {1\over 4}[1 + m(s_i + s_j) + r_{\alpha}s_is_j]$,
or
\begin{eqnarray}
\rho_{\alpha}(\pm\pm) = {1\over 4}(1 \pm 2m + r_{\alpha}) \geqslant 0,
\hspace{1cm}\rho_{\alpha}(+-) = \rho_{\alpha}(-+) = {1\over 4}(1 - r_{\alpha}) \geqslant 0. \label{prob3}
\end{eqnarray}

Here $m = \langle s_i \rangle$ is a magnetization, $r_{\alpha} =
\langle s_is_j \rangle_{i,j\in\alpha}$ is a pair-correlation function
of the cluster $\alpha$, and these five parameters span a 5-dimensional
space. For the $1:2$ structure $m = r_1 = r_4 = -1/3$, $r_2 = 1/3$, $r_3 =
0$, and two configuration probabilities are zero, namely $\rho_1(++)$
and $\rho_4(++)$. This accounts for 2 faces of the 5-dimensional
space. If there are no other constraints on the
$m,r_{\alpha}$ ($\alpha=1,{\ldots},4$), the structure cannot be a vertex. 
But there are other constraints. The point is, that $\rho \geqslant 0$ are
sufficient conditions only if all $\nu_{\beta}$ are defined on the same
cluster. This is not the case here and neither it was in
Ref~\cite{mcco53}. There are in general additional relations between the
probabilities for configurations defined on different clusters, since one can imagine 
these clusters being parts of a single larger cluster. 
Let's take a simple example. Consider
the 4NN pair ($\alpha=4$) and mark its sites with $(+)$ if
$s_i=+1$, $(-)$ if $s_i=-1$, and $(\cdot)$ for the site in the middle
(to make it different from the NN pair). Also consider
for a moment a linear triplet which contains a 4NN pair and two NN
pairs. Since $\rho(s_i,s_j)=\sum_{s_k=\pm1}\rho(s_i,s_j,s_k)$ and all
probabilities are non-negative, it is obvious that
$\rho(+\cdot-) = \rho(++-) + \rho(+--) \leqslant \rho(++) + \rho(--)$,
which using (\ref{prob3}) gives a relation
\begin{eqnarray}
1 - r_4 \leqslant 2 + 2r_1.  \label{newconstr}
\end{eqnarray}
This relation defines a new constraint and a new hyperplane, which is
not equivalent to any of the constraints in (\ref{prob3}). It is not
that simple to obtain all the constraints. In general one has to
consider clusters large enough to contain {\em all} the interaction
clusters (e.g. a double-cube), then to project the resulting CP on the
space of $m, r_{\alpha}$ ($\alpha=1,{\ldots},4$), and see if the
vertex coresponding to the desired structure survives. Fortunately,
here we can use the result~\cite{mcco53} that the cube+4NN pair
interaction cluster is missing one constraint for the 1:2 phase to be a
vertex. Since the constraint (\ref{newconstr}), which is not
equivalent to those already considered in Ref~\cite{mcco53}, applies
here also, and the corresponding hyperplane includes the 1:2 phase, we
conclude that the 1:2 structure is a possible ground state.

\begin{figure}
\caption{Schematic view of ordered phases mentioned in the text. 
The structures are projected onto the (001) plane.
Open circles indicate a sequence of B'' atoms along the [001] direction.
Half-filled black-white (white-black) circles represent B', B'' 
(B'', B') atoms alternating along the [001] direction.}
\label{fig0}
\end{figure}

\begin{figure}
\caption{ Phase diagrams $(T,x)$ (a) and $(T, \mu)$ (b)  obtained by
the CVM with the interaction constants $v_2=0.2$, $v_3=0.06$ and
$v_4=0$.  Solid line - 8+7 - point approximation ($v_1=1$), dashed line
-- 27-point approximation ($v_1=\infty$). The two-phase regions, marked by
numbers are: 1 -  disordered+1/4 and 2 - 1:1+3/8}
\label{fig1}
\end{figure}

\begin{figure}
\caption{
Phase $(T,x)$ (a) and $(T, \mu)$ (b) diagrams obtained by
the CVM with the interaction constants $v_2=0.2$, $v_3=0.06$, and
$v_4=0.02$. Solid and dashed lines - same as in Fig.\ref{fig1}, 
dotted line -- 8+7-point approximation with the variational space,
reduced to one- and two-point correlation functions only. Monte Carlo
results are shown by circles. Phases to the right of the 3/8' phase
and to the left of the 1/4 phase are not shown. The two-phase regions, 
marked by numbers are: 1 - disordered+1/4 and 2 - disordered+1:2}
\label{fig2}
\end{figure}

\begin{figure}
\caption{
The metastability limits of 2:1 and 1:1 phases at $\mu=2.1$ 
obtained with the CVM (solid lines) and Monte Carlo calculations (circles).
The energy $E = \langle {\cal H} \rangle$ per site is presented for
$v_2=0.2$, $v_3=0.06$, and $v_4=0.02$. The dashed lines are CVM free-energies
which intersect at the first-order phase transitions (pointed to by an arrow).}
\label{fig3}
\end{figure}

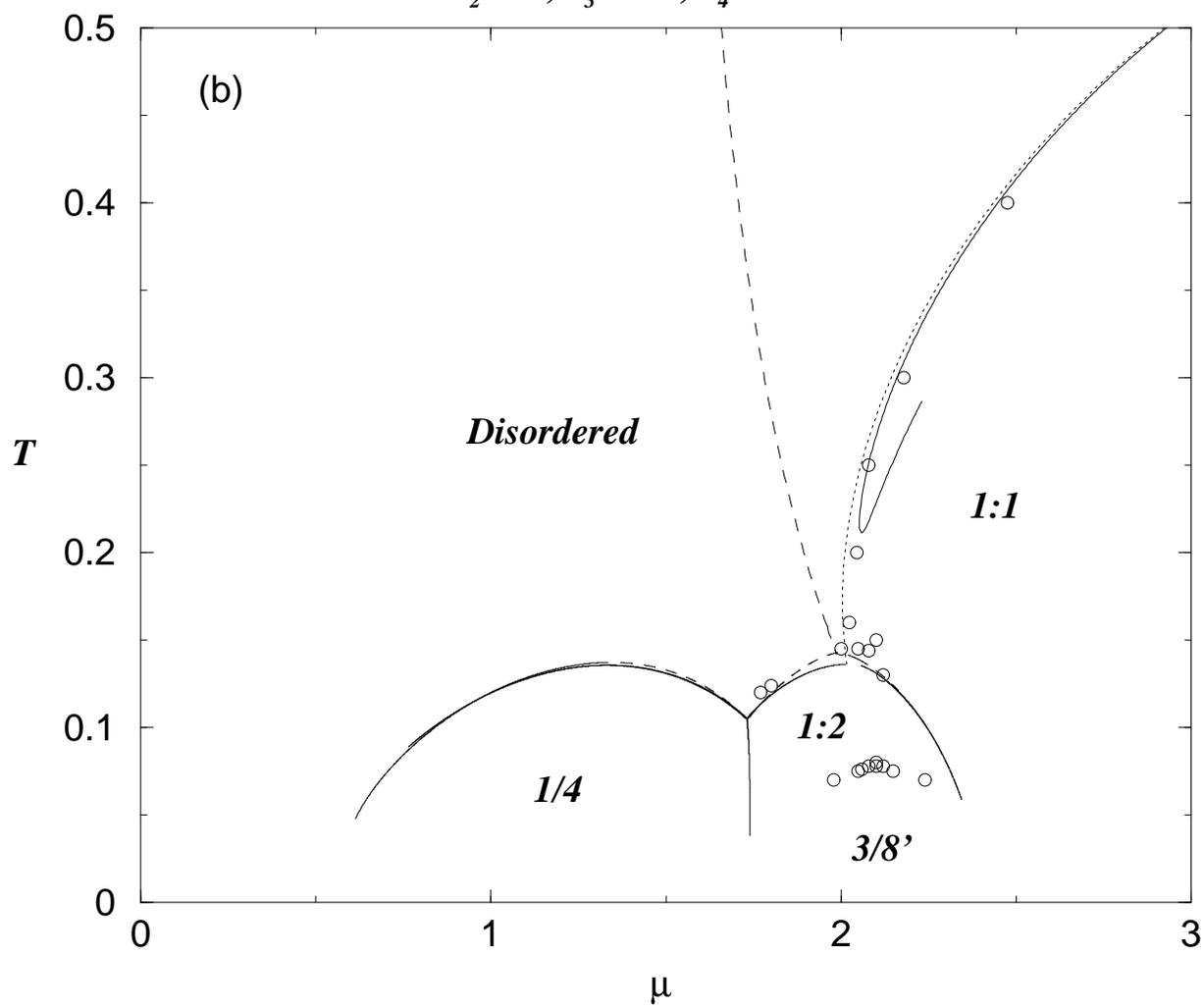
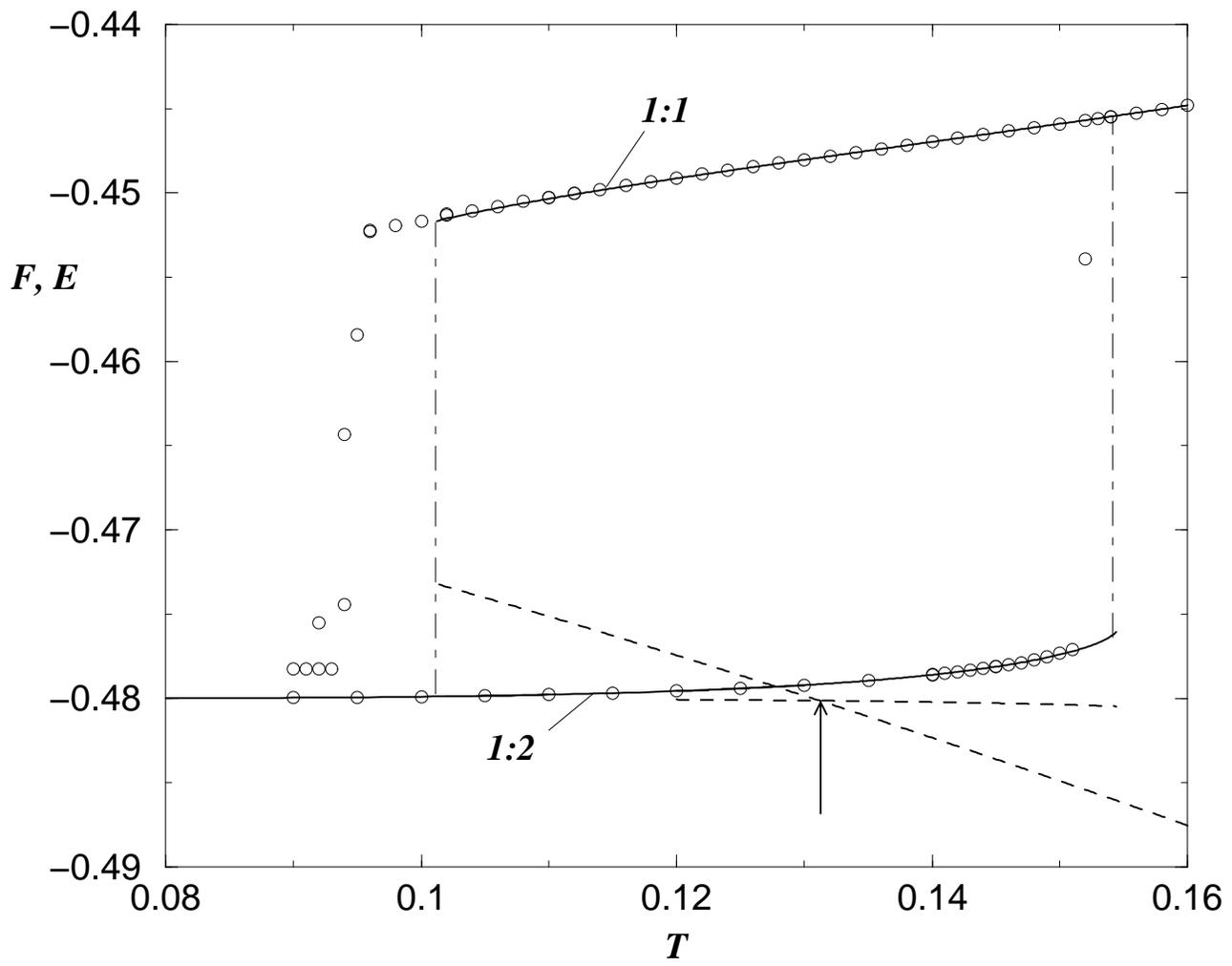
\end{document}